\documentclass[aps,prl,twocolumn,superscriptaddress,showpacs]{revtex4}

\usepackage{graphicx}
\usepackage{amssymb}
\usepackage{amsmath}
\usepackage{bm}
\usepackage{subfigure}
\usepackage{graphpap}
\usepackage{multirow}
\usepackage{notes2bib}
\usepackage[version=3]{mhchem} 

\begin{document}
\title{\textit{Ab initio} Simulation of Optical Limiting: The Case of Metal-Free Phthalocyanine}
\author{Caterina \surname{Cocchi}}
\affiliation{Centro S3, CNR-Istituto Nanoscienze, Via Campi 213A, I-41125 Modena, Italy.}
\affiliation{Humboldt-Universit\"at zu Berlin, Institut f\"ur Physik und IRIS Adlershof, D-12489 Berlin, Germany}
\email{caterina.cocchi@physik.hu-berlin.de}
\author{Deborah \surname{Prezzi}}
\affiliation{Centro S3, CNR-Istituto Nanoscienze, Via Campi 213A, I-41125 Modena, Italy.}
\author{Alice \surname{Ruini}}
\affiliation{Centro S3, CNR-Istituto Nanoscienze, Via Campi 213A, I-41125 Modena, Italy.}
\affiliation{Dipartimento di Scienze Fisiche, Informatiche e Matematiche, Universit\`a di Modena e Reggio Emilia, I-41125 Modena, Italy}
\author{Elisa \surname{Molinari}}
\affiliation{Centro S3, CNR-Istituto Nanoscienze, Via Campi 213A, I-41125 Modena, Italy.}
\affiliation{Dipartimento di Scienze Fisiche, Informatiche e Matematiche, Universit\`a di Modena e Reggio Emilia, I-41125 Modena, Italy}
\author{Carlo A. \surname{Rozzi}}
\affiliation{Centro S3, CNR-Istituto Nanoscienze, Via Campi 213A, I-41125 Modena, Italy.}
\date{\today}
\begin{abstract}
We present a fully \textit{ab initio}, non-perturbative description of the optical limiting properties of a metal-free phthalocyanine, by simulating the effects of a broadband electric field of increasing intensity.
The results confirm reverse saturable absorption as leading mechanism for optical limiting phenomena in this system and reveal that a number of dipole-forbidden excitations are populated by excited-state absorption, at more intense external fields. 
The excellent agreement with the experimental data supports our approach as a powerful tool to predict optical limiting, in view of applications.
\end{abstract}
\maketitle
Optical limiting (OL), a strong attenuation of the light transmittance through a material at increasing input intensity, is a prototype nonlinear phenomenon of relevance for an entire class of devices \cite{sun-rigg99irpc}.
It has received rising attention in view of the need to protect light-sensitive elements, including the human eye, from optical damage \textit{e.g.} due to novel intense and tunable light sources \cite{tutt-bogg93pqe}.
Organic compounds are good candidates as OL materials \cite{span99jmc,he+08cr}, and among them macrocyclic molecules are particularly effective, due to their large and delocalized $\pi$-electron network \cite{dini+03ssi}.
Phthalocyanines and porphyrins present several advantages in this field \cite{calv+04sm,dela+04cr}: in addition to their excellent chemical processability and versatility, they are characterized by a complex spectrum in the UV -- visible range, which can be tuned by chemical modifications, either by decorating the organic network or by inserting a metal atom \cite{perr+94optl,dela+98jmc,ofla+03am}.
A number of experimental studies in the last decades provided a general picture of the main physical mechanisms driving optical limiting in these molecules \cite{wei+92apb,perr+96sci,hana+01ccr}.
Previous theoretical studies (see \textit{e.g.} Ref. \cite{li-li12oc}), however, are mostly limited to a phenomenological level of description, and are therefore unable to clarify the microscopic origin of OL phenomena and to predict the behavior of different systems.

Within the framework of real-time time-dependent density functional theory (TDDFT), we perform a fully \textit{ab initio} investigation of the OL properties of a metal-free phthalocyanine.
The calculated optical spectra are the key ingredient to determine the OL curve in the frequency region of interest, upon increasing irradiance.
A large excited-state absorption around the green visible region emerges as the driving mechanism for reverse saturable absorption (RSA) and hence for OL effects in these systems.
Our results reproduce the nonlinear behavior of the intensity-dependent absorption coefficient, which is a fingerprint of optical limiting, and confirms the validity of our approach to characterize the response of macromolecules to strong electromagnetic fields.
The insight into the microscopic origin of these phenomena, provided by \textit{ab initio} calculations, represents an added value to predict optical limiting in new materials.
Our proposed \textit{ab initio} approach also provides a bridge to the existing empirical models, since it allows us to check their limits of applicability.

\begin{figure}
\centering
\includegraphics[width=.48\textwidth]{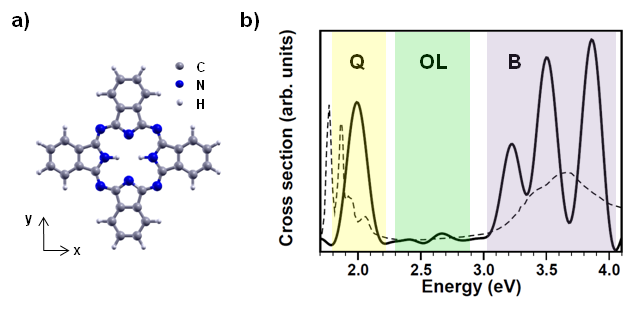}%
\caption{(Color online) \textbf{a)} Ball-and-stick representation of a metal-free phthalocyanine with the central pyrrole rings bound to two H atoms (\ce{H2}-Pc).
\textbf{b)} Optical absorption spectrum of \ce{H2}-Pc, computed in linear regime (solid line, $F_{in}$=0.001 J/cm$^2$), in agreement with the experimental spectrum (dashed line, from Ref. \cite{dela+04cr}). The visible ($\sim$ 2 eV, Q-band) and the near UV region ($\sim$ 3-4 eV, B-band) are separated by a large absorption valley (2.3 -- 2.9 eV, OL).
}
\label{fig1}
\end{figure}
%

As a prototype system, we study a free-base phthalocyanine (Pc) \ce{C32H18N8}, with two H atoms bonded to the central pyrrole rings (\ce{H2}-Pc, see Fig. \ref{fig1}a).
This molecule, which is the basic organic backbone for a manifold of macrocyclic complexes, presents itself interesting OL characteristics \cite{qu+03cpl,hong+05ml,math+07oc,venk+08apb,kuma+08spie}.
The linear optical properties of \ce{H2}-Pc have been studied within a number of theoretical approaches, ranging from model Hamiltonians \cite{orti+90jcp,cort+03jpca}, to semi-empirical \cite{henr-sund72tca} and \textit{ab initio} configuration-interaction methods \cite{toyo+97jpca}, as well as linear-response TDDFT \cite{zhon+10jmgm,yost+11jpcc}.
Optical experiments \cite{edwa-gout70jms,dela+04cr} (see Fig. \ref{fig1}b, dashed line) reveal that the spectrum of \ce{H2}-Pc is characterized by two main absorption bands, one in the UV (B-band) and one in the visible region (Q-band). 
The broad transparent window in between is of particular interest for OL properties \cite{dela+04cr}.
The absorption increase upon intense illumination in this band can conveniently protect the human eye from intense visible radiation.

The calculation of the optical absorption spectrum of \ce{H2}-Pc (solid line, Fig. \ref{fig1}b) is performed by real-time propagation of the electron density \cite{yaba-bert96prb}, as implemented in the \texttt{octopus} code \cite{cast+06pssb, note-xc}.
The optical properties of the system are obtained through a Fourier transform of the molecular dipole induced by an impulsive electric field $\mathbf{E}$, corresponding to a broad-band excitation.
The external field is represented by a suitable sudden change in the phase of the Kohn-Sham orbitals \cite{note-kick}.
For a more direct comparison with the experiments, in the remainder of the paper we refer to the fluence $F$ of the incoming radiation, instead of the field amplitude. 
The fluence corresponds to the field intensity $I$ multiplied by the time-step of the propagation $\delta t$: $F$ = $I \delta t$ = $\dfrac{c n \epsilon_0}{2} |\mathbf{E}|^2$.
In our simulations $\delta t \simeq 10^{-18}$ s and the time propagation is carried out for about 23 fs.
For arbitrarily small values of the input fluence ($F_{in}$), the system responds linearly to the external perturbation. 
To probe the nonlinear regime, $F_{in}$ spans three orders of magnitude, ranging from 0.001 to 5.816 J/cm$^2$, as in typical experiments.
From the time propagation of the electron density \cite{ullr11book,marq+12book}, the frequency-dependent dynamical polarizability and hence the absorption spectrum are computed.
\begin{figure}
\centering
\includegraphics[width=.48\textwidth]{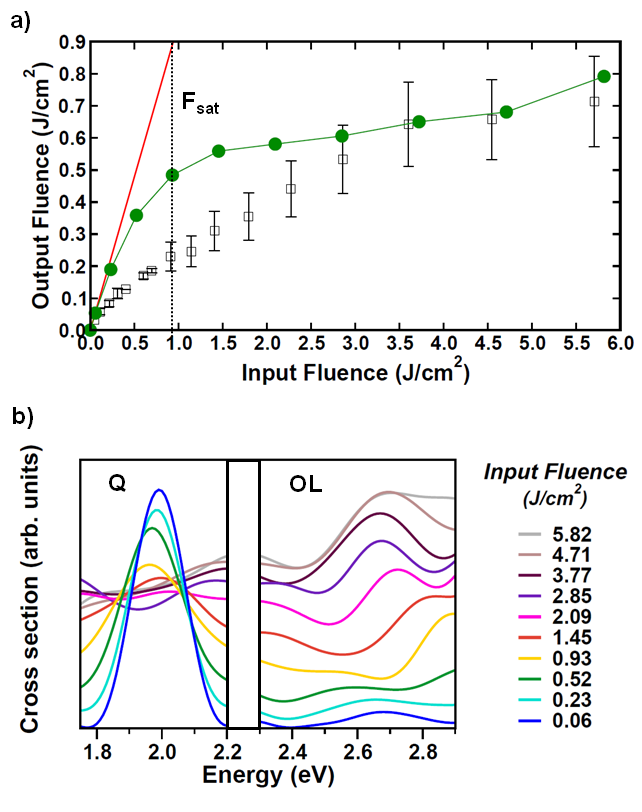}%
\caption{(Color online) \textbf{a)} Optical limiting curve (filled dots) of \ce{H2}-Pc, obtained by integrating the absorption cross section of the TDDFT spectra in the OL region (see Fig. \ref{fig1}b). The experimental results (empty squares, Ref. \cite{venk+08apb}) for an analogous molecule in solution, are reported for comparison. The solid red line indicates the linear regime, with $\alpha_0$ = 0.43 cm$^{-1}$, computed from the absorption cross section under the lowest input fluence ($F_{in}$=0.001 J/cm$^2$, see spectrum in Fig. \ref{fig1}b, solid line). The saturation fluence, $F_{sat}$, indicates the threshold between linear and nonlinear regime.
\textbf{b)} Nonlinear TDDFT spectra of \ce{H2}-Pc calculated at increasing external fields.
}
\label{fig2}
\end{figure}

In Fig. \ref{fig2}a we present the optical limiting curve of \ce{H2}-Pc (filled dots).
It is computed from the TDDFT spectra at the adiabatic local density approximation level \cite{note-xc, note-spectra}, at increasing input fluence (see Fig. \ref{fig2}b), by integrating the photo-absorption cross-section ($\sigma$) over the OL window (2.3 -- 2.9 eV). 
The integrated $\sigma$ is the key quantity derived from our simulations which is physically linked to optical limiting. 
Though, it cannot be directly related to experiments, where one measures both $F_{in}$ and $F_{out}$, the latter being the relevant observable for OL.
In experimental conditions there are also additional environmental parameters (e.g. the molar density $N$ and the sample thickness $L$), which affect the actual value of $F_{out}$ but are not physically connected to optical limiting, as confirmed by the observation of this phenomenon in different sample embedding \cite{hong+05ml, math+07oc}.
Optical limiting is indeed a property of the active absorber and not of the solvent.
Under this assumption, it is possible to relate $\sigma$, obtained from the simulations, with the experimentally observed $F_{out}$, through the evaluation of the absorption coefficient $\alpha$.
Known the sample thickness $L$, $\alpha$ is extracted from experimental data through the Beer-Lambert law (assuming sample reflectance uniformly equal to zero) $F_{out} = F_{in} e^{-\alpha L }$ \cite{note-fluence}.
Given the molar density $N$, $\alpha$ is obtained from calculations via the relation $\alpha =N \cdot \sigma$.
Hence, to quantitatively compare with experimental results, we need to substitute the sample thickness and the molar density with values directly taken from experiments. 
Here we refer to Ref. \cite{venk+08apb} and thus we assign $L$=1 mm and $N=10^{-4}$ mol.
The OL curve obtained from the experiment of Venkatram \textit{et al.} of a \ce{H2}-Pc in solution \cite{venk+08apb} is also shown in Fig. \ref{fig2}a (empty squares).
For sufficiently weak perturbations, the absorption coefficient $\alpha$ is a constant, independent of the field intensity \cite{note-lin}.
In this regime, $F_{out}$ is directly proportional to $F_{in}$ (solid line, Fig. \ref{fig2}a): the absorption is described by the Beer-Lambert law for $\alpha=\alpha_0=$ 0.43 cm$^{-1}$, where $\alpha_0$ is obtained for the lowest input fluence $F_{in}$=0.001 J/cm$^2$ (see spectrum in Fig. \ref{fig1}b).
As $F_{in}$ grows, the transmitted fluence deviates from linearity and the curve assumes the characteristic shape of optical limiting.
For $F_{in}$ exceeding the saturation fluence $F_{sat}$ -- defined as the value of the input fluence at which the output is one half of its value in linear regime \cite{tutt-bogg93pqe} (dashed line in Fig. \ref{fig2}a) -- $F_{out}$ still grows, but much more slowly than in linear regime. 
In summary, TDDFT results are correctly reproducing the trend and the overall shape of the OL fingerprint. 
They are also in excellent quantitative agreement with experimental values, especially in the two extreme limits in which the incoming radiation fluence is well below and well above threshold. 
This confirms the effectiveness of our approach to accurately describe the OL effect.

The analysis of the optical features in the TDDFT spectra offers a deeper insight into the mechanisms responsible for optical limiting and represents the actual added value of the theoretical approach adopted here.
In excellent agreement with the available experimental data \cite{dela+04cr}, the linear optical spectrum of \ce{H2}-Pc (see Fig. \ref{fig1}b) is characterized by intense peaks in the visible (Q-band) and UV region (B-band) and by an absorbing valley in-between (OL), which represents the region of interest for optical limiting.
The analysis of the linear-response spectra \cite{casi95book} supports us in the interpretation of the main optical features of \ce{H2}-Pc.
The Q-band ($\sim$ 2 eV) is composed of two excitations, $Q_x$ and $Q_y$, separated in energy by a few hundreds meV, and polarized along the $x$ and $y$ axis of the molecules, respectively (see Fig. \ref{fig1}a).
They stem from transitions between the HOMO ($A_u$ symmetry) and a doublet of quasi degenerate states ($B_{3g}$ and $B_{2g}$ symmetry) forming the LUMO.
The absorption intensity of $Q_x$ and $Q_y$ is given by the large transition dipole between the involved single-particle states.
In the UV region, between 3 and 4 eV, the B-band is composed of a manifold of peaks, due to dipole-allowed excitations polarized in the ($x$, $y$) plane, which involve deeper occupied and higher empty states, with alternating $g$ and $u$ character.
The absorbing valley between the Q- and the B-band is characterized by a number of dark excitations, related to the promotion of electrons from the highest occupied to the lowest empty orbitals having the same parity.
While in linear regime these dipole-forbidden excitations cannot contribute to the optical absorption, strong external fields allow the population of these states.
In fact the cross section in the OL region grows significantly at increasing values of the input fluence (see Fig. \ref{fig2}b, right panel), giving rise to optical limiting effects, as shown in Fig. \ref{fig2}a.
A different behavior appears instead for the Q-band (Fig. \ref{fig2}b, left panel): at increasing input fluence the peak broadens and it is washed out for $F_{in} > F_{sat}$.
However its integrated spectral weight and peak location do not undergo significant modifications.

\begin{figure}
\centering
\includegraphics[width=.45\textwidth]{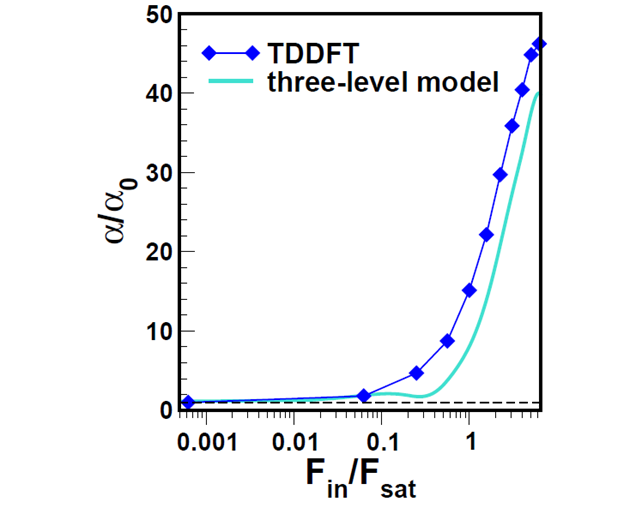}%
\caption{(Color online) Normalized absorption coefficient $\alpha/\alpha_0$, as a function of the normalized input fluence $F_{in}/F_{sat}$. The diamonds indicate the values of $\alpha$ computed directly from the TDDFT absorption cross section, while the solid curve is obtained analytically from the three-level model in Eq. \ref{eq:alpha_model}. The constant value $\alpha/\alpha_0$ = 1, corresponding to the ideal linear case, is inserted for comparison (dashed line).}
\label{fig3}
\end{figure}

Our approach, based on \textit{ab initio} calculations, allows us to validate the few-state models \cite{tutt-bogg93pqe,ofla+03am}, typically adopted to interpret experimental results. 
They are based on the assumption that, under intense external fields, excited-state absorption involves a limited number of levels above the ground state (usually three or five altogether).
The lifetime of these metastable states is assumed to be long enough compared to the light pulse, such that they can be populated and hence contribute to the absorption.
As a consequence of excited-state absorption, the absorption coefficient $\alpha$ increases with respect to the intensity (or equivalently the fluence) of the incident field.
In the three-level model, as derived and explained in details in Ref. \cite{ofla+03am}, the dependence of $\alpha$ on the input fluence is analytically described by the expression
\begin{equation}
\alpha = \dfrac{\alpha_0}{1+\dfrac{F_{in}}{F_{sat}}} \left(1+ \kappa \dfrac{F_{in}}{F_{sat}}\right).
\label{eq:alpha_model}
\end{equation}
This equation in turns depends on a few parameters, whose values are usually obtained through the fit with experimental data, namely the weak field absorption coefficient $\alpha_0$, the saturation fluence $F_{sat}$ and the ratio of the excited-state to the linear cross section $\kappa=\sigma / \sigma_0 \gg 1$.
To check the validity of this model, we compare in Fig. \ref{fig3} the behavior of $\alpha$, computed directly from TDDFT cross sections (diamonds) without making use of Eq. \ref{eq:alpha_model}, with the values obtained from Eq. \ref{eq:alpha_model}, in which $\alpha_0$, $F_{sat}$ and $\kappa$, calculated from first principles, are plugged in (solid line).
In the limit of weak input fluence ($F_{in} \ll F_{sat}$), the absorption coefficient is independent of the fluence ($\alpha = \alpha_0$, dashed line in Fig. \ref{fig3}).
As $F_{in}$ increases, $\alpha / \alpha_0$ grows steeply, up to about a factor 50, when $F_{in}$ exceeds $F_{sat}$ by about one order of magnitude.
It is worth noting that the curves in Fig. \ref{fig2} are obtained by considering the OL window between 2.3 and 2.9 eV.
By focusing on a narrower band around the green frequency (\textit{e.g.} 2.3 -- 2.5 eV), $\alpha / \alpha_0$ increases up to over two orders of magnitude in the explored fluence range.
On the other hand, by integrating over a large portion of the UV-vis spectrum of the molecule (0.0 -- 5.0 eV), the absorption coefficient exceeds $\alpha_0$ only of a factor 3, thus showing the relevance of the OL properties of \ce{H2}-Pc in the visible region around the green band.
The consistent trend of the absorption coefficient computed from TDDFT and the three-level model (Eq. \ref{eq:alpha_model}) confirms that excited-state absorption involves only a limited number of excited states, which are responsible for RSA and OL effects in the considered system. 

In analogous way, \textit{ab initio} calculations can clarify the role of other nonlinear optical phenomena, such as two-photon absorption (TPA), in the OL behavior of \ce{H2}-Pc.
In presence of TPA, the absorption law reads \cite{suth+03book}:
\begin{equation}
I_{out} = \dfrac{I_{in} e^{-\alpha L}}{1+\beta L_{eff} I_{in}}.
\label{eq:TPA}
\end{equation}
This expression corresponds to the integrated Beer-Lambert equation, damped by the denominator, which depends explicitly on the input intensity $I_{in}$, on the effective thickness of the sample $L_{eff} = (1-e^{- \alpha_0 L})/\alpha_0$ and on the TPA absorption coefficient $\beta$.
It is evident already from the differential form of the Beer-Lambert equation $dI/dz = -\alpha I - \beta I^2$, that $\beta$ provides a second order contribution to the absorption, in terms of incident intensity.
In the case of \ce{H2}-Pc, the values of $\beta$ computed from TDDFT results by inverting Eq. \ref{eq:TPA} are several order of magnitudes lower compared to $\alpha$ ($\beta \sim$10$^{-7}$ -- 10$^{-8}$ cm/W) \cite{note-beta}.
The agreement with the typical experimental values of $\beta$ \cite{tutt-bogg93pqe,venk+08apb} confirms the inefficacy of TPA alone to drive OL effects in phthalocyanines.
The efforts to improve $\beta$, and hence TPA, in different classes of materials represent however an important branch of study in nonlinear optics (see for review Ref. \cite{dela+04cr}).

In conclusion, we provided a fully \textit{ab initio} description of optical limiting effects in a metal-free phthalocyanine, considered as a prototypical system for macrocyclic molecules and for a wide range of compounds.
The employed methodology is non-perturbative in essence, and therefore does not require the analysis of linear and quadratic response theory coefficients.
The linear and nonlinear regimes are described on the same footing by a time-propagation of the Kohn-Sham system, parametrically dependent on the input light intensity.
The potential of TDDFT to study optical limiting is clearly demonstrated by our results, which are in quantitative agreement with the available experimental data and validate the few-level models, typically adopted to fit them.
From the analysis of the TDDFT spectra, we observed that upon intense illumination excited-state absorption promotes the population of dipole-forbidden excitations in the visible region, which contribute to the absorption strength.
The presence of a manifold of dark excitations in the linear regime therefore appears as an important requirement for the manifestation of OL behavior.
Our approach allows theoretical understanding of the relevant microscopic mechanisms leading to optical limiting, and opens the way to a rational design and systematic prediction of this nonlinear effect.

The authors are grateful to Nicola Spallanzani at CINECA for computational support.
This work was supported by Fondazione Cassa di Risparmio di Modena (project COLDandFEW), by the Italian Ministry of Education, University and Research (FLASHit project, Grant No. FIRB-RBFR12SW0J), by the PRACE Research Infrastructure (Grant nr. 1562) and by the European Community (CRONOS project, Grant No. 280879).
CPU time was granted by CINECA through the ISCRA projects nr. HP10CG3OGS
(OCT12), HP10C6XSZH (CARONTE) and HP10CYY9C5 (OPLA).



\begin{thebibliography}{10}
\bibitem{note-xc}
The system is described on a grid with spacing 0.18 \AA{} on a set of spheres
  of radius 8 \AA{} centered around each atom. We adopt the local density
  approximation, using a modified version of the Perdew-Zunger correlation
  potential \cite{perd-zung81prb}. We checked that the spectra do not change
  significantly by adopting generalized gradient approximation (GGA-PBE) for
  the exchange-correlation functional. The core-electrons are treated via
  Troullier-Martins pseudopotentials \cite{trou-mart91prb}.
\bibitem{note-fluence}
In this form the Beer-Lambert law can be equivalently written in terms of the
  intensity or of the fluence.
\bibitem{note-lin}
In Ref. \cite{venk+08apb} the transmitted fluence is assumed to be 70$\%$ of
  the input one in conditions of linear absorption. From our TDDFT
  calculations, $F_{out}$ is 96$\%$ of $F_{in}$ in linear regime. This explains
  the different behavior between our results and the experimental data in the
  transient region between linear and nonlinear absorption regime.
\bibitem{note-beta}
In the calculation of $\beta$ from Eq. \ref{eq:TPA}, we consider a laser pulse
  with a duration of 6 ns, as from typical experimental values
  \cite{venk+08apb}. The corresponding intensities are compatible with the
  values of fluence adopted in our TDDFT simulations.
\end{thebibliography}
\end{document}